# Nanolithographic templates using diblock copolymer films on chemically heterogeneous substrates

Vassilios Kapaklis[1,*], Spyridon Grammatikopoulos[2], Roman Sordan[3], Alessio Miranda[3], Floriano Traversi[3], Hans von Känel[3], Dimitrios Trachylis[2], Panagiotis Poulopoulos[4] and Constantin Politis[2]

[1] *Department of Physics and Materials Science, Uppsala University, Box 530, S-751 21 Uppsala, Sweden*
[2] *Engineering Science Department, School of Engineering, University of Patras, 26504 Patras, Greece*
[3] *Department of Physics, Milan Polytechnic, Via Anzani 42, 22100 Como, Italy*
[4] *Materials Science Department, School of Natural Sciences, University of Patras, 26504 Patras, Greece*

**Abstract**

The orientation of the lamellae formed by the phase separation of symmetric diblock copolymer thin films is strongly affected by the wetting properties of the polymer blocks with respect to the substrate. On bare silicon wafers the lamellae of polystyrene-b-polymethylmethacrylate thin films tend to order parallel to the wafer surface, with the polymethylmethacrylate block preferentially wetting silicon. We have developed a methodology for inducing the arrangement of lamellae perpendicular to the substrate by using chemically modified substrates. This is done by chemisorbing a self-assembled monolayer of thiol-terminated alkane chains on thin gold films deposited on silicon wafers. We also show that it is possible to spatially control the perpendicular orientation of the lamellae at sub-micron length scales by using simple chemical patterns and etch them, in order to produce nanolithographic templates. This method may be of great technological interest for the preparation of well-defined templates using block copolymer thin films.

**Keywords**: block copolymer, block copolymer lithography, e-beam lithography, self-assembled monolayers, atomic force microscopy

**PACS**: 83.80.Uv, 85.40.Hp, 81.16.Dn, 87.64.Dz

---

[*] Corresponding author. Email: vassilios.kapaklis@fysik.uu.se. Tel: +46 18 471 3627.

# 1. Introduction

Self-assembly is a universal phenomenon[1,2], occurring everywhere in nature and can lead to technologically interesting approaches, for the bottom-up fabrication of functional nanostructures[3]. Especially for the case of self-assembly in molecules the interest in the last decades has grown exponentially. The structures emerging from molecular self-assembly are of great interest for basic research and technological applications, since the processes that generate them may be used in nanoscale engineering and architecture[4,5,6].

Copolymers or heteropolymers are polymers derived from two or more monomeric species. Block copolymers are made up of blocks of different polymerized monomers. The latter are quite large molecules, which can be tailored concerning their lengths and chemical properties and present a novel approach for the fabrication of nanoscale templates. Self-organization in such molecules arises due to the phase separation of the distinctive blocks that make up the block copolymer. The interplay of entropy of chain folding and the connectivity of the polymer blocks results in morphologies, which are mainly controlled by the volume ratio of the blocks[7].

Lamellar morphologies and the self-organized patterns observed in such systems may be of great technological interest. Selective etching and removal of one block, after stable lamellar morphologies have been achieved, can present a new simple, parallel and cost-effective method for the production of nanoscale templates[8]. It is thus important to develop methodologies for the successful control of the development of such morphologies. A key parameter affecting these morphologies and the arrangement of the lamellae with respect to the substrate is the wetting property of each polymer block to the substrate and air interface of the block copolymer film[9]. By balancing the wetting of the polymer blocks to the substrate it is possible to drive the phase separation to form lamellae oriented perpendicular to the substrate. Using silicon wafers as substrates to deposit thin block copolymer films, does not ensure this perpendicular arrangement. This is due to the preferential wetting of one block to the silicon oxide ($SiO_x$) surface on the wafer[10]. In this case the lamellae orient parallel to the film surface exposing the other block to the film-air interface when the film thickness is in multiples of half integer $L_0$ values (where $L_0$ is the lamellar microdomain periodicity in the bulk)[11]. Here we show that in a polystyrene-b-polymethylmethacrylate (PS-b-PMMA) diblock copolymer, in which PMMA preferentially wets the $SiO_x$ substrate, it is possible to induce a perpendicular arrangement of lamellae by chemical modifications of the substrate.

Our approach aims to control the wetting of the polymer blocks to the substrate. It involves the use of a self-assembled monolayer (SAM) of functionalized alkane chains, *Y(CH$_2$)$_n$X* , where *Y* and *X* are end groups[12]. Previous attempts using SAMs of silanes chemisorbed on silicon substrates have been successful in inducing epitaxial self assembly of the phase separated domains in block copolymer thin films[13].

More specifically, in the case of thiols (*HS(CH$_2$)$_n$X*) one of the end groups is the *HS* group that chemisorbs onto gold or platinum substrates[14]. The absorption process produces a self-assembled molecular layer of thiols in an ordered fashion[15]. The surface energy of the monolayer can be easily tuned by modifying the head group *X*. In our case the head group is hydroxyl (*-OH)*.

Based on the pioneering work of *Heier et al.*[16], in which they investigated the influence of chemically modified substrates on the structure and phase separation of thin diblock copolymer films, we demonstrate that it is possible to spatially control the arrangement of the lamellae relative to the substrate at the sub-micrometer scale. For this purpose we have used pre-patterned substrates of thin gold films, on which we chemisorb self-assembled monolayers of thiols. Our results show that a perpendicular arrangement takes place over the sites where gold has been deposited and thiols have been chemisorbed sequentially. Furthermore, we were able to etch the PMMA lamellae at these sites, using oxygen plasma, leaving the rest of the film unharmed. This fact may be of great technological interest for the spatial confinement of nanolithographic templates produced by the use of block copolymer thin films[17,18,19,20].

## 2. Experimental section

### 2.1. Large area and patterned Ti/Au film deposition

Ti(2 nm)/Au(10 nm) films were patterned by e-beam lithography using a standard lift-off process. The films were deposited on Si substrates by e-beam evaporation.

### 2.2. SAM preparation

SAMs of 11-Mercapto-1-undecanol 97 % (MUD) were chemisorbed on Pt. Samples were immersed in a Ethanol/Tetrahydrofurane solution containing the MUD molecules (60 g Ethanol, 20 g Tetrahydrofurane and 250 mg MUD). Immersion time was kept in all cases at 15 min. All chemicals were purchased from Aldrich and used as received.

### 2.3. Block copolymer

PS-b-PMMA diblock copolymer was purchased from PSS Mainz, Germany. In our studies we used a symmetric copolymer (50 % PS – 50 % PMMA) with total molecular weight of 84.4 kDa (42.2 kDa PS, 42.2 kDa PMMA). The poly-dispersivity index (PDI) for this molecule is 1.06 and the lamellar microdomain periodicity in bulk is approximately 42 nm ($L_0$). Thin films of this block copolymer were spin coated from a toluene solution (1.5 % wt.) at 4000 rpm and resulted in a film thickness of

~21 nm (0.5 $L_0$) for horizontal oriented lamellar films. Film thickness was measured by cross section scanning electron microscopy (SEM) and atomic force microscopy.

*2.4. Temperature annealing*

In order to improve the phase separation morphology and contrast after spin coating of the block copolymer thin films, samples were thermally annealed in vacuum at 195 °C in a horizontal tubular furnace. This temperature is near a "softening" point for this copolymer and facilitates the rearrangement and ordering of the blocks in the phase-separated structure. Annealing was carried out for 2 h in all cases.

*2.5. Atomic Force Microscopy (AFM)*

The phase separation morphology and topography in our films was observed by AFM, using a Nanosurf Mobile S system. All images were recorded in tapping mode, using silicon cantilevers oscillating at frequencies of 190 kHz (force constant 48 N/m). Phase sensitive detection of the cantilever bending, yields a measure for the tip indentation and is determined by the local elasticity of the surface of the sample[21]. In this way we can resolve the different phase regions of PS and PMMA in our phase-separated block-copolymer films, through their different elastic properties.

*2.6. Plasma etching*

The samples were etched by $O_2$ plasma in a plasma asher (30 s, 500 ml/min, 1000 W).

**3. Results and discussion**

In Figure 1 we present a schematic of the way in which the SAMs of MUD grow on the Au film, which is deposited onto silicon. Measurements of the static contact angle of a water droplet (50 μl) onto a bare silicon wafer ($SiO_x$), a gold film, as well as MUD treated film, showed that the contact angle changes drastically. In our experiments, the MUD surface was more hydrophilic than the native silicon oxide or gold, exhibiting small contact angles (typically <10°). In previous studies it has been shown that PS interacts preferentially with gold substrates but dewets silicon oxide surfaces[22,23]. The increase of the substrate surface energy, by applying the SAM, is the cause for the increase of hydrophilicity[16].

Due to the latter, we expect more balanced wetting properties for the copolymer blocks on the substrates treated with MUD which could result in a perpendicular alignment of the phase separated lamellae. In order to investigate this, we used substrates fully covered by a gold film and then treated with MUD SAMs and measured their properties with respect to the lamellar morphology, obtained in the phase separation of the block copolymer films. Figure 2 shows an AFM scan of such a sample. After spin coating the copolymer solution and thermal annealing at 195 °C, we record images similar to Figure 2 on the whole surface of the film. The film thickness measured by AFM and SEM was found to be 20 nm. The total sample size is ~ 1×1 cm$^2$, which compared to the features in the phase-separated structure, is extremely large. Defects mainly occur due to dust particles, which accumulate on the substrate, if not working in clean room conditions, and also due to impurities in the solution used for spin coating. It is evident that the phase separation structure results in a lamellar periodicity $L_0$, of ~ 42 nm, which is expected for this copolymer and is exhibited in thin block copolymer films when the arrangement on the lamellae is perpendicular[11].

We extend this approach for inducing a perpendicular lamellar arrangement by patterning the gold films. Patterned gold lines were produced by means of e-beam lithography and a lift-off technique. We tested various line widths, which all are below the micron range. After applying a SAM of MUD on the patterned sample, spin coating of the copolymer solution and thermal annealing, we recorded AFM images such as those shown in Figure 3. In Figure 3 (a) a large area scan of such a sample is shown, which contains parallel lines of various widths. The lower right corner of this scan contains lines with widths well below 100 nm. Figure 3 (b) shows a line with 800 nm width. A small topographic signal is recorded in an area that resembles a line. On either side this line the film surface is flat, without forming any kind of holes or islands, indicating that the copolymer lamellae probably orient parallel to the silicon surface. This observation verifies once more the accurate control of the film thickness, which is 0.5 $L_0$. The gold lines with the chemisorbed MUD SAM are only an extremely small part of the total sample surface, which is mainly dominated by the native $SiO_x$ of the Si substrate. As already mentioned PMMA preferentially wets the polar $SiO_x$ surface of the substrate, whereas PS has a low surface energy and wets the vacuum interface, resulting in an asymmetric wetting case[16]. For the latter, only half-integer values of $L_0$ result in flat films[24]. Indeed, this is confirmed by the phase image recorded for exactly the same area and clearly depicts that perpendicular alignment has taken place only over the gold line. On the sample surface next to the patterned line no phase contrast can be observed, indicating that the air interface of the film is wetted only by one of the blocks, which in our case is PS. The region with perpendicular alignment exhibits very small width variations along the underlying gold line. An interesting for applications consequence of this structure, is that by selectively etching PMMA it is possible to develop the copolymer film pattern onto the substrate, only over the gold line sites[25]. The rest of the film, which is covered by the PS lamellae on the top, will not be affected as much.

Subsequently, structures like these were etched using oxygen plasma, in order to remove PMMA from the copolymer film and leave behind a structure consisting of the remaining PS. Figure 4 (a) shows a similar to Figure 3 (b), 800 nm wide line that has been etched and a typical scan line profile. As can be seen, the film has been etched over the line pattern and a clearly stronger topographic signal is recorded there, indicating that PMMA has been etched away. The rest of the film looks unharmed, verifying that the top layer of the film is PS. This acts as a protective cover and does not allow the etching of the underling film. Figures 4 (b) & (c) show more line patterns of smaller widths after plasma etching. The method is successful in etching even the thinnest lines with an approximate width of 50 nm.

Summarizing all the experimental data on our samples, the proposed structure for all the studied films is depicted in Figure 5. Figure 5 (a) shows a typical section analysis for a thin PS-b-PMMA film of thickness 0.5 $L_0$, spin-coated and annealed on a silicon wafer. The asymmetric wetting of the interfaces observed in such a film, results in PMMA preferentially wetting the silicon surface while PS aligns on top, wetting the air interface. Figure 5 (b) shows the structure of a PS-b-PMMA film spin coated on a MUD SAM on gold, where both PS and PMMA wet the substrate and produce a phase separated structure, where the lamellae arrange themselves perpendicular to the substrate. Even though the film thickness was maintained at 0.5 $L_0$, the microdomain periodicity in these films is double, because of the arrangement of the copolymer blocks within these domains, as shown on the magnification schematic. Viewed from the top, such a striped structure may produce an image as the one seen in Figure 2. The morphologies observed in Figure 3 (b) are a combined result of the previous two structures. This is shown schematically in Figure 5 (c), where perpendicular alignment takes place over the MUD covered gold lines. Finally, in Figure 5 (d) the structure of a plasma-etched film is shown, in which PMMA has been removed and PS is left behind, forming a nanolithographic template. In all cases the perpendicular aligned lamellae have a self-assembled striped structure and do not show any preferred orientation with respect to the gold line axis. If the chemical modification pattern of the substrates was to commensurate with the lamellar periodicity $L_0$ of the block copolymer, then it might be possible to observe epitaxial self-assembly of the block domains[13].

## 4. Conclusions

We have demonstrated that it is possible to induce a perfect perpendicular lamellar arrangement of symmetrically phase-separated PS-b-PMMA block copolymer films, by applying functionalized thiol SAMs on gold thin films deposited on silicon substrates. The resulting films have a flat topography but show a strong contrast in phase imaging, due to the phase-separated structure. Using the very different wetting properties of the block copolymer thin film with respect to the bare silicon substrate and the SAM treated gold film; we have successfully confined the induced perpendicular arrangement only to the sites where we have deposited gold line patterns. We have also shown that it is possible to etch these patterns, removing PMMA and producing nanolithographic templates with features of several nanometers. It will also be important to investigate in future studies, whether it is possible to further induce an orientation to the self-organized striped pattern formed by the block copolymer lamellae over the gold sites. This approach may lead to new insights in inducing patterns to the phase separation of block copolymer thin films, which could be of great technological interest in the near future.

**Figure captions**

**Fig. 1**. Schematic showing the arrangement during chemisorption, of thiol alkane chains onto a gold surface.

**Fig. 2.** Tapping mode AFM scan of a phase-separated PS-b-PMMA film spin-coated on a MUD-treated substrate. Defect-free perpendicular lamellar arrangement can be observed. The image on the left corresponds to the topography of the sample, which is flat. The image on the right is the phase scan of the AFM instrument for the exactly same sample area. Two distinct phases can be indentified on the sample surface corresponding to the phase separated PMMA and PS blocks of the copolymer.

**Fig. 3**. **(a)** Tapping mode AFM scan of a phase separated PS-b-PMMA film spin coated over a sample containing patterned gold lines of various line widths, with chemisorbed MUD SAMs. The film is defect free and flat with small topographic variations over the patterned gold lines. **(b)** Magnifying such a line in order to resolve the film structure confirms the perpendicular phase separation of the copolymer film over the gold line, **(c)** as clearly shown in the phase scan. The line width in this image is 800 nm. Overall, the perpendicular arrangement of the lamellae is well contained within the width of the line.

**Fig. 4** **(a)** AFM scan of a plasma etched phase separated PS-b-PMMA film, over an 800 nm wide patterned gold line covered by a SAM of MUD. The topography of the image indicates that PMMA has been etched away, leaving behind PS and a self organized pattern similar to that of Figure 3 (b). The arrow on the right side of the image indicates the position of the line scan depicted on the right. Etched line patterns of **(b)** 400 nm and **(c)** 50 nm widths.

**Fig. 5**. **(a)** Phase-separated film structure of a symmetric PS-b-PMMA copolymer on a silicon wafer. The structure shows the case of asymmetric wetting and film thickness of 0.5 $L_0$, which is necessary to obtain flat films. **(b)** Film structure of the same copolymer over a gold thin film covered by a MUD SAM. The magnification depicts the arrangement of the copolymer block within the domains, resulting in a lamellar microdomain periodicity of $L_0$. **(c)** Proposed structure for a film like the one shown in Figure 3 (b), and **(d)** the same film that has been plasma etched and PMMA has been removed.

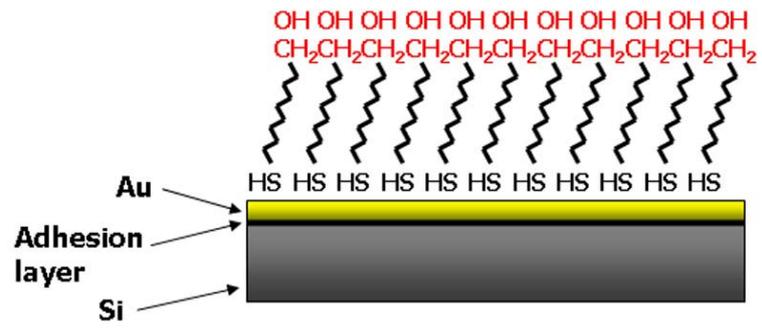

**Fig. 1. V. Kapaklis et al.**

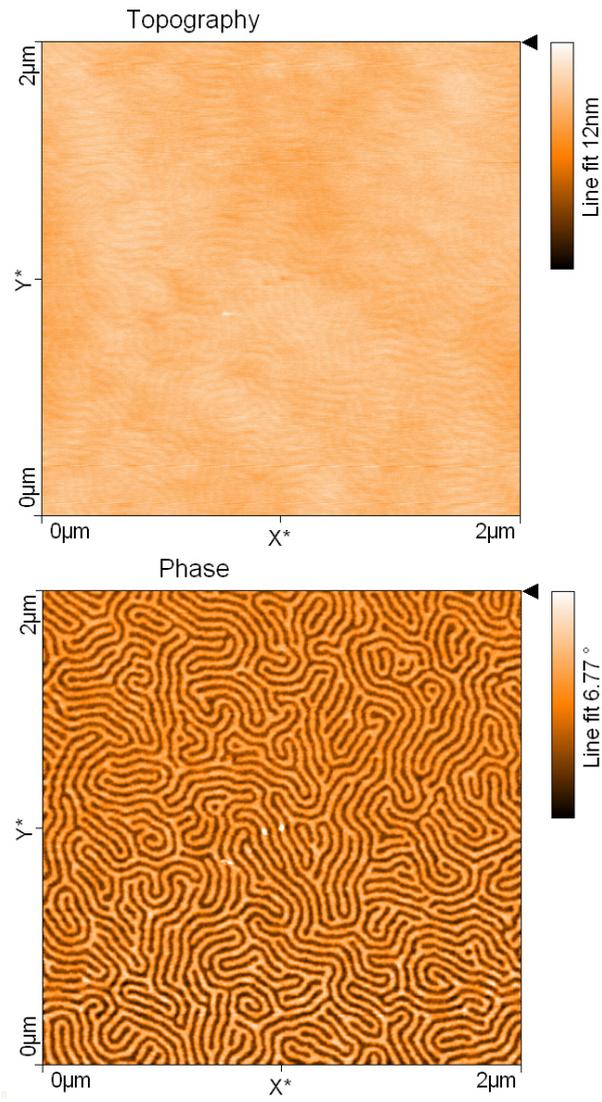

**Fig. 2. V. Kapaklis et al.**

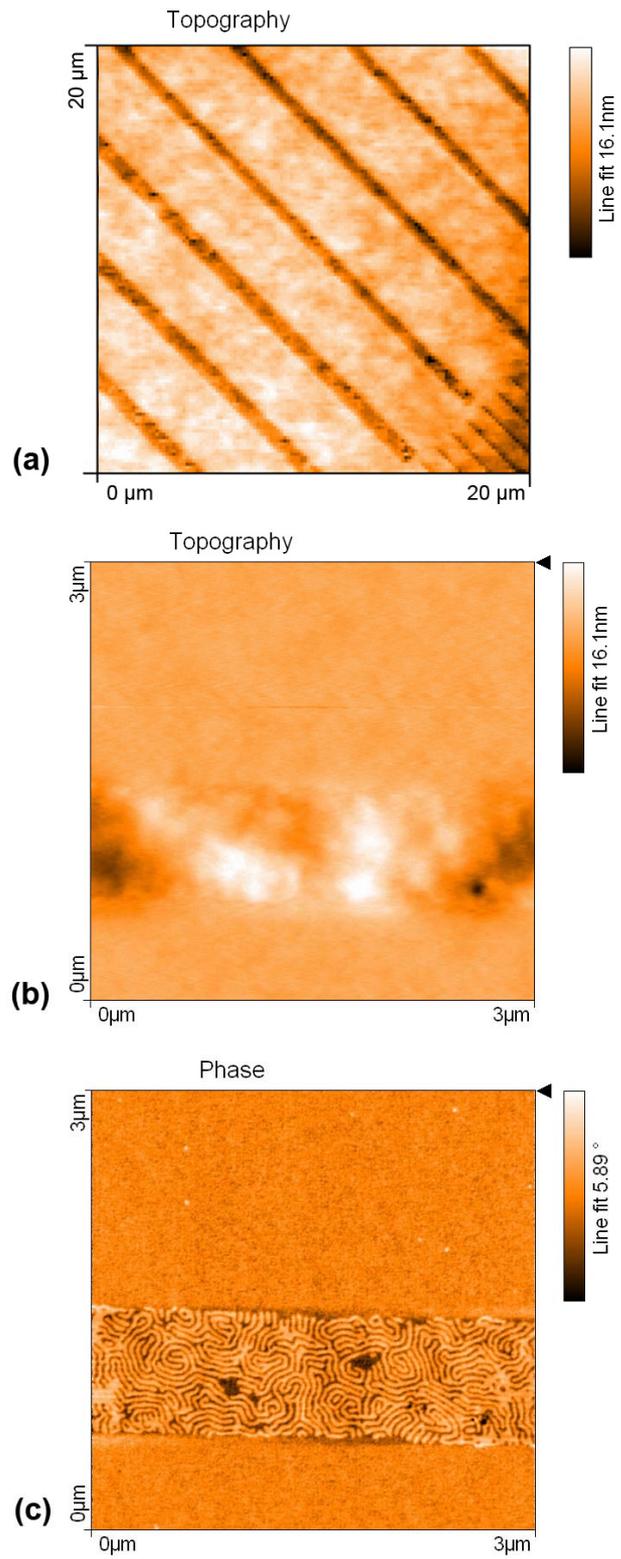

Fig. 3. V. Kapaklis et al.

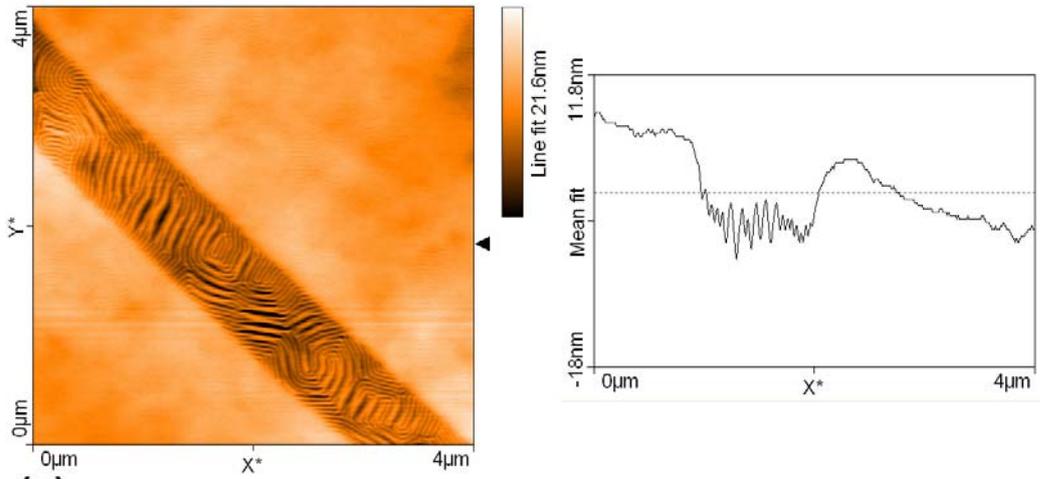

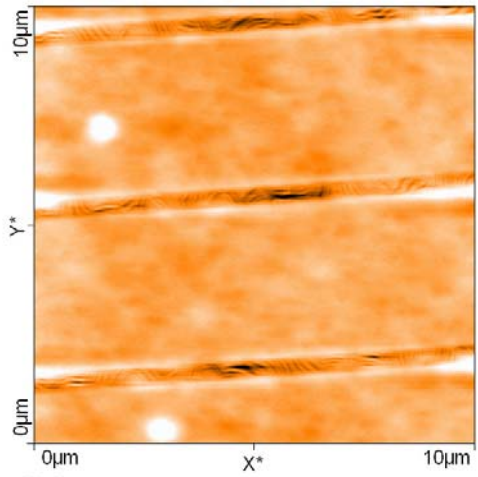
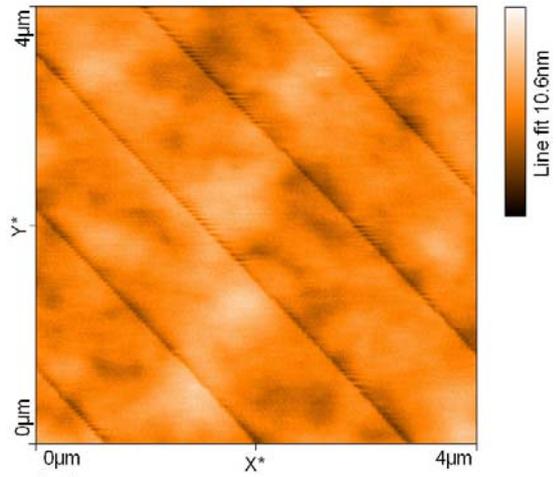

**Fig. 4. V. Kapaklis et al.**

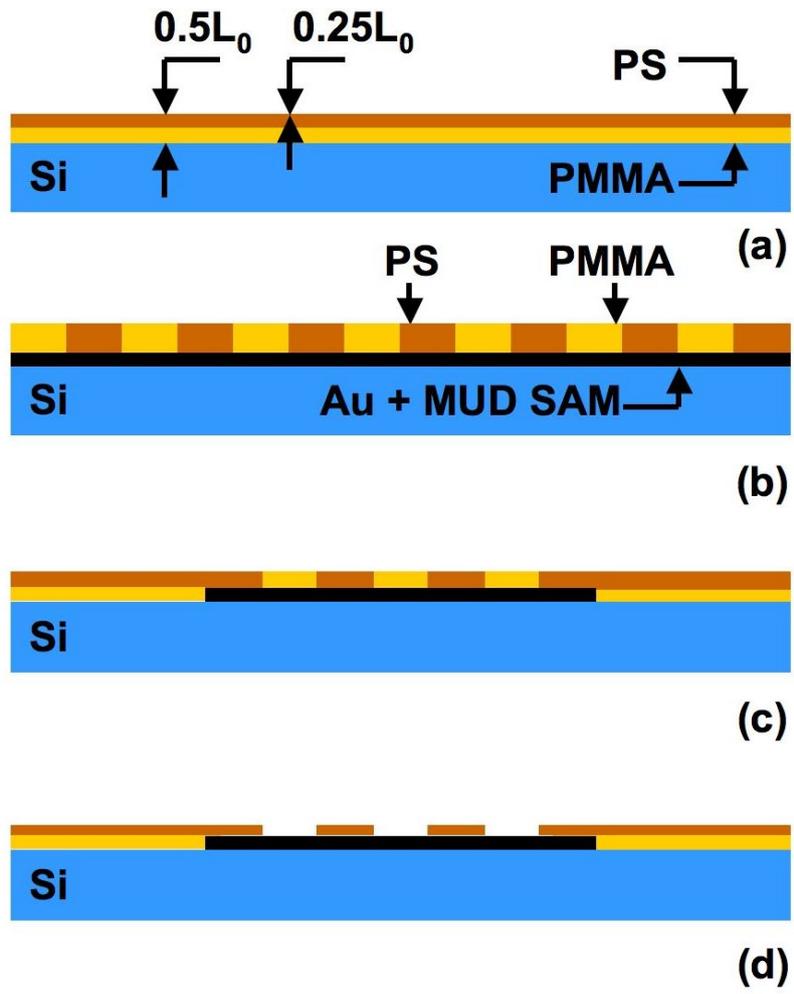

Fig. 5. V. Kapaklis et al.